\documentclass[journal=jpcafh,manuscript=article]{achemso}

\usepackage[english]{babel}
\usepackage{natbib}
\usepackage[mathlines]{lineno}
\usepackage[T1]{fontenc}
\usepackage{xcolor}
\usepackage{multirow}
\usepackage{rotating}

\author{Jorge Vargas\footnote{Current address: Unidad Academica de Fisica, Universidad Autonoma de Zacatecas,
Calz. Solidaridad Esq. Paseo de la Bufa S N, Zacatecas, Mexico. C.P. 98060}}
\affiliation{Department of Physics, The University of Texas at El Paso, El Paso, Texas, 79968}
\author{Peter Ufondu}
\affiliation{Department of Physics, The University of Texas at El Paso, El Paso, Texas, 79968}
\author{Tunna Baruah}
\affiliation{Department of Physics, The University of Texas at El Paso, El Paso, Texas, 79968}
\altaffiliation{ Computational Science Program, The University of Texas at El Paso, El Paso, Texas, 79968}
\affiliation{Department of Physics, The University of Texas at El Paso, El Paso, Texas, 79968}
\author{Yoh Yamamoto}
\affiliation{Department of Physics, The University of Texas at El Paso, El Paso, Texas, 79968}
\author{Koblar A. Jackson}
\affiliation{Physics Department and Science of Advanced Materials Program, Central Michigan University, Mt. Pleasant Michigan, 48859}
\author{Rajendra R. Zope}
\affiliation{Department of Physics, The University of Texas at El Paso, El Paso, Texas, 79968}
\altaffiliation{ Computational Science Program, The University of Texas at El Paso, El Paso, Texas, 79968}
\email{rzope@utep.edu}

\title{Importance of self-interaction-error removal in density functional calculations on water cluster anions}

\keywords{Self-interaction correction, water cluster anions, FLOSIC}

\begin{document}

\begin{abstract}
 Accurate description of the excess charge in water cluster anions 
 is challenging for standard semi-local and (global) hybrid density functional 
 approximations (DFAs).
 Using the 
 recent unitary invariant implementation of the Perdew-Zunger self-interaction 
 correction (SIC) method using Fermi-L\"owdin orbitals, we assess the effect of self-interaction
 error on the vertical detachment energies of  
 water clusters anions with the local spin density approximation (LSDA),
 Perdew-Burke-Ernzerhof (PBE)  generalized gradient approximation, and the strongly constrained and
 appropriately normed (SCAN) meta-GGA functionals. Our results show that 
 for the relative energies of isomers with respect to reference 
 CCSD(T) values, the uncorrected SCAN functional has the smallest deviation of 
 21 meV, better than that for the MP2 method. The performance of SIC-SCAN 
 is comparable to that of MP2 and is better than SIC-LSDA and SIC-PBE, but
 it reverses the ordering of the two lowest isomers for water 
 hexamer anions. 
 Removing self interaction error (SIE) corrects the tendency of LSDA, PBE, and SCAN to over-bind the extra electron.
 The vertical detachment energies (VDEs) of water cluster anions, obtained 
 from the total energy differences of corresponding anion and neutral clusters, are significantly improved by removing self-interaction and are better than the hybrid B3LYP functional,
 but fall short of MP2 accuracy. Removing SIE results in substantial 
 improvement in the position of the eigenvalue of the extra electron. The negative of the
 highest occupied eigenvalue after SIC provides an excellent approximation to 
 the VDE, especially for SIC-PBE where the mean absolute error with respect to CCSD(T) is only 17 meV, the best among all approximations compared in this work.
\end{abstract}

\section{Introduction} 
The hydrated electron is a system that has long attracted the attention of the scientific 
community\cite{Hart69,Garret05,kulkarni2000ab,Pathak08,Herbert15, Herbert17} 
due to its importance in chemical and 
biological processes, such as atmospheric chemistry\cite{Gross11} or radiation damage in DNA,
\cite{Alizadeh15} to name just two.
In this context, systematic study of water cluster anions can be useful for 
obtaining insights into the behavior and evolution of electron hydration. 
Indeed, since the first observation of a free hydrated electron in the gas phase\cite{Armbruster81}, 
a large number of studies have been performed on water cluster anions 
resulting in debate about whether the extra electron is bound in a delocalized surface state or bound internally in a cavity.\cite{Hammer05,Turi05,Jacobson11,Rossky88,Turi12,Wilhelm19}
Even before its first observation, theoretical models for the hydrated electron were proposed\cite{Kevan80,Kevan81}.
Later, improvements in experimental techniques allowed direct comparison with 
theoretical results. Photoelectron spectroscopy
\cite{Coe90,Kim96,Kim98,Shin04,Verlet05,Coe08,Ma09} 
and vibrational spectroscopy 
\cite{Bailey96,Ayotte99,Diken04,Asmis07,Guasco10} 
techniques have been used to garner information about the 
structural and electronic properties of water cluster anions. These 
experiments have provided data for the vertical detachment energy (VDE) of
small water clusters. The VDE is the energy required to 
remove an excess electron from an anion. Moreover, a combination of simulations and experimental 
vibrational spectra has been used to identify the structure of stable 
isomers\cite{Ayotte99,Hammer05a,Hammer05b}. 
It is, however, in general difficult to assign a particular 
isomer to the observed experimental spectra as the experiments 
usually sample non-equilibrium ensembles of clusters. A detailed description
of the challenges involved in ascribing specific spectroscopic features to an
individual isomer of (H$_2$O)$_6^-$ is provided by Choi and Jordan in Ref.
\citenum{Choi09}.

 Accurate description of a hydrated electron poses a significant challenge 
 to density functional approximations (DFAs). In general, DFAs are inadequate for describing 
the binding of the excess electron. As a result, most 
computational studies on small anionic water clusters 
\cite{Hammer05a,Hammer05,Lee96,Kim96,Tsurusawa99,Kim99,Weigend99,
Lee03,Lee05,Herbert05,Herbert06,Yagi08,Unal18,Zho18}
have used a post Hartree-Fock (HF) method like the M{\o}ller-Plesset
perturbation method (MP2) or the coupled cluster method with single, double and perturbative triple excitations CCSD(T).  CCSD(T) is computationally very
demanding and has only been used to study small clusters and often in
these studies, structural relaxation is carried out by a faster
method like MP2. 

 Many failures of the DFAs have been ascribed to the self-interaction error (SIE) present in 
the approximate exchange-correlation functionals. The SIE occurs due to the incomplete cancellation of the self-Coulomb by the approximate self-exchange energy. This error is particularly dominant in the local and semi-local
approximations and is mitigated to some extent in the hybrid DFAs due to the
addition of HF exchange. It has, however, been found that 
the most popular global hybrid functional B3LYP \cite{Becke93} significantly overestimates the VDEs of anionic water clusters. 
On the other hand, a recently proposed non-empirical meta-GGA functional (SCAN)\cite{SCAN15} has been found 
to provide an excellent description of the structural, electronic, and 
dynamic properties of liquid water\cite{Chen17}. 
The SCAN meta-GGA functional, unlike most other DFAs, 
also predicts the energetics of gas-phase water hexamers 
and ice phases with quantitative accuracy\cite{SCAN15}. 

In this work
we examine the role of SIE in three non-empirical
functionals that belong to the lowest three rungs of Jacob's ladder
on the VDEs
of water. The functionals used here are the local density approximation (LDA), generalized gradient approximation (GGA) given by Perdew, Burke, Ernzerhof (PBE) \cite{PBE}, and the SCAN meta-GGA functional \cite{SCAN15}. 
The incorrect asymptotic form of the DFA potential caused by SIE 
is expected to have a particularly strong impact on the description of anionic systems which typically have a weakly bound extra electron in a diffuse orbital. We explictly remove SIE using the self-interaction correction (SIC) applied to the DFA's.  We study both the effect of SIC on the orbital energy of the highest occupied electron orbital and on the total energy difference between corresponding anion and neutral systems.  
We also check the error made by using the SCAN functional for calculating the VDE.  
The details of the self-interaction correction method and the computation scheme is presented in the next section followed by the results and discussion.

\section{Methodology}
In 1981, Perdew and Zunger proposed an orbital-wise correction  to
remove the self-interaction error \cite{PZSIC81} in density 
functional approximations. In the Perdew-Zunger self-interaction correction (PZSIC) method 
the exchange-correlation  energy is corrected as 
\begin{equation}\label{ESIC}
	E_\mathrm{XC}^\mathrm{SIC}[\rho_\uparrow,\rho_\downarrow] = 
	-\sum_{i,\sigma}^{N_\sigma} \{U[\rho_{i\sigma}] +E_\mathrm{XC}[\rho_{i\sigma},0]\},
\end{equation}
where  $i$ runs over the $N_\sigma$ occupied
orbitals of spin $\sigma$, and $\rho_{i\sigma}$ 
is the $i^{th}$ orbital density. The terms
$U[\rho_{i\sigma}]$ and $E_\mathrm{XC}[\rho_{i\sigma},0]$ are the
exact self-Coulomb and approximate self exchange correlation (XC) energies, respectively.
The correction vanishes when $E_\mathrm{XC}$ is the exact XC 
functional.  Pederson et al.  have shown that 
the orbitals minimizing the PZ-SIC total energy must satisfy the conditions known as the localization equations:\cite{Lin84,Lin85}
\begin{equation}\label{LE}
	\langle \phi_{j\sigma} |V_{j\sigma}^\mathrm{SIC}
	-V_{i\sigma}^\mathrm{SIC}| \phi_{i\sigma} \rangle,
\end{equation}
where $V_{i\sigma}^\mathrm{SIC}$ is the SIC potential for the i$^{th}$ orbital.  Satisfying the localization equations is a computationally slow process and the self-interaction corrected energy obtained as a result is not guaranteed to be size-consistent.

A recent scheme for SIC proposed by Pederson,
Ruszinzsky and Perdew \cite{FLOSIC14}
circumvents the need for satisfying Eq.\ [\ref{LE}]. 
The localized orthonormal set of orbitals is derived from Fermi orbitals which depend on the density matrix and
the spin density as:
\begin{equation}
	\phi_{i\sigma}^\mathrm{FO}(\mathbf{r}) = \frac{\sum_j \psi_{j\sigma}^*(\mathbf{a}_{i\sigma})
	\psi_{j\sigma}(\mathbf{r})} {\sqrt{\sum_j |\psi_{j\sigma}(\mathbf{a}_{i\sigma})|^2}}
	= \frac{\gamma_\sigma(\mathbf{a}_{i\sigma},\mathbf{r})} {\sqrt{\rho_\sigma(\mathbf{a}_{i\sigma})}},
\end{equation}
where $\gamma_\sigma(\mathbf{a}_{i\sigma},\mathbf{r})$ is the single-particle
density matrix of the KS system, and $\mathbf{a}_{i\sigma}$ are a set of points in real space called the Fermi
orbital descriptors (FODs). The Fermi orbitals are normalized, but are not othogonal.  They are orthogonalized using
L\"owdin's method of symmetric orthonormalization \cite{Lowdin50}
resulting in an orthonormal set of local orbitals called  Fermi-L\"owdin orbitals (FLOs). The positions of the FODs  determine the Fermi orbitals and different choices lead to different total energies. The optimal positions of the FODs are obtained in a procedure that is analogous to a molecular geometry optimization. The gradients of the energy with respect to the FODs can be calculated\cite{Pederson15,PEDERSON_AAMOP} and used in a pre-conditioned limited-memory Broyden, Goldfarb, Shanno (LBFGS) algorithm\cite{liu1989limited}. The FODs are updated  after each self-consistent FLOSIC calculation. The  optimization is carried out until the forces on all  FODs  drop below 0.0001 Ha/Bohr.

The Fermi-L\"owdin orbital based self-interaction correction (FLO-SIC)  method is implemented 
in the FLOSIC code\cite{yamamoto2019fermi,FLOSICCodep} that is based 
on the UTEP-NRLMOL code \cite{Jackson90,Pederson00}.
This code uses a Gaussian basis set \cite{Porezag99},
and a variational integration mesh \cite{Pederson90}
to perform numerically precise calculations on molecules composed of non-relativistic atoms. 
The FLOSIC implementation has been used to study a number of different properties for systems ranging in size from atoms \cite{Kao17b,withanage2018question} and small molecules to larger molecules such as Mg-porphyrin and C$_{60}$ \cite{Pederson16}.  FLOSIC has been used 
to study various properties ranging from energetic properties such as atomization energies\cite{yamamoto2019fermi}, barrier heights\cite{sharkas2018shrinking,Shahi19, yamamoto2019fermi}, magnetic properties\cite{kao2017role,joshi2018fermi} to density dependent properties such as dipole moment\cite{johnson2019effect} and polarizability\cite{withanage2019self}. 
   
We use the default Pederson-Porezag NRLMOL basis \cite{Porezag99} that is specially optimized 
for the  Perdew-Burke-Ernzerhof 
(PBE) \cite{PBE} generalized gradient approximation (GGA). The calculations are spin-polarized for systems with net spin.
To obtain an accurate description of water anions,  extra diffuse functions are added to account for the more diffuse charge distribution
in these systems \cite{Herbert05}.
We used the same exponents as used by Yagi {\it et al.} \cite{Yagi08}. We have verified that these exponents give converged results for the VDEs.
The exponents are $9.87\times 10^{-3}$ au, $8.57\times10^{-3}$ au,
and $3.72\times10^{-3}$ au for oxygen $s$, $p$, and hydrogen $s$ respectively.

Yagi and coworkers reported the anionic water cluster geometries optimized at the MP2 level which they used subsequently to perform the CCSD(T) calculations \cite{Yagi08}. 
To facilitate a direct comparison with the earlier MP2 and CCSD(T) results,
we used the same set of MP2 optimized geometries from Ref. \citenum{Yagi08} 
in our calculations. 

Each water molecule has five electrons of each spin. 
The FODs representing the valence electrons of each molecule form 
a tetrahedral structure with the center at the oxygen atom and two of the vertices along the two O-H bonds. The tetrahedral structure can be seen in Fig. \ref{dimer}. The FOD representing the oxygen core orbital is found to be at the oxygen nuclear position. The extra FOD for the anions is initially placed at the central region of the singly occupied molecular orbital obtained at the PBE level. The position of this FOD is then optimized along with all the others using the pre-conditioned LBFGS routine outlined above.

The VDE can be calculated as the total energy difference
between the energy of the anion and neutral cluster at the geometry of the anion. The negative of the energy of the highest occupied molecular orbital (HOMO) also mimics the negative of the electron removal or detachment 
energy\cite{perdew1982density,levy1984exact,perdew1997comment,PhysRevB.60.4545}.
The HOMO eigenvalue for an anion in a DFA calculation is generally found to be positive, corresponding to an unbound outer electron. Therefore 
the detachment energies can only be calculated from total energy differences when using DFAs. As shown below, the anion HOMO in FLOSIC-DFA calculations is negative and a good approximation to the removal energy.  We examine the VDEs calculated from total energies as well as from the HOMO eigenvalues for the FLOSIC calculations.

\section{Results and discussion}
The anionic water clusters contain a weakly bound extra electron. 
The 20 water isomers $(\mathrm{H}_2\mathrm{O})_n^-$ in the range $n=2-6$
studied here can be grouped by various types of extra electron binding motifs.
The clusters studied here and other slightly different clusters have been reported in several studies
with different names \cite{Kim96,Tsurusawa99,Lee03,Hammer05,Unal18}.
For direct comparison with the work of Yagi and coworkers\cite{Yagi08}, we follow
their  naming conventions. The anionic water clusters are classified as linear(L), double acceptor (AA), donor(D), and internal (I).
In linear (L) clusters
the water molecules are bound by successive hydrogen bonds (HB). 
These structures tend have the smallest number of HBs among all the isomers. 
The AA type clusters have one double-acceptor water molecule such that its hydrogen atoms not involved in any HB. In the D type clusters the extra electron is bound collectively by dangling O-H bonds. In I type anions the extra electron is trapped internally, as in a cavity.

\begin{figure}
    \centering
    \includegraphics[width=0.4\textwidth]{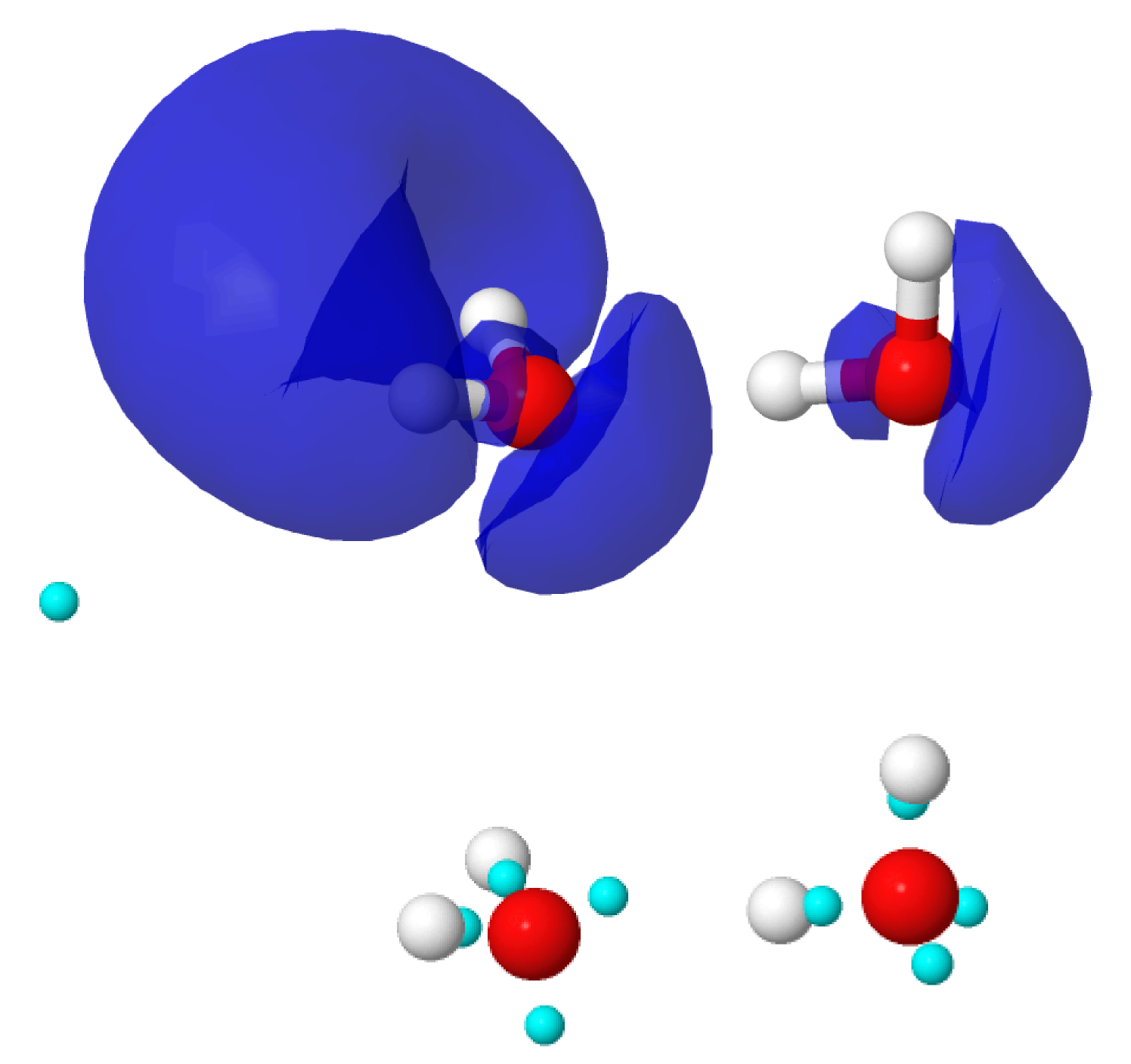}
    \caption{({\bf Top})Electron density difference (blue) of anionic water dimer anion and ({\bf bottom}) the optimized FODs shown as cyan spheres.}
    \label{dimer}
\end{figure}

The water clusters studied here are presented in figures 1--5, where the positive electron density difference (EDD) between the anionic and the neutral systems shows the charge density of the extra electron. The density difference also contains rearrangements of the neutral molecule density due to the presence of the extra electron, but such differences are insignificant compared to density of the extra electron itself. The FLOSIC calculations of the water anions were started with the same FOD positions as for the neutral water cluster but with one extra FOD. All the FODs are fully re-optimized for the anionic clusters. For simplicity, only the optimized extra FOD 
position is shown in Figs 2-5 and an image including all the FODs is shown for the anionic water dimer in Fig. \ref{dimer}.

 The number and strength of the HBs in a cluster determine its stability\cite{Yagi08}. For most of the water cluster sizes studied here, the isomers with dangling O-H bonds (D) are the lowest energy structures. The D structures are cyclic, with every water molecule
being both donor and acceptor of a hydrogen bond and the dangling O-H bonds oriented 
toward the same direction where most of the extra charge is accommodated \cite{Yagi08}. Moreover, it was shown that the orientations of the dangling O-H bonds in most stable anionic clusters are different from those for the neutral clusters \cite{Yagi08}. In neutral water clusters of size up to n=5, the most stable isomers have the dangling OH bonds on alternating sides of a quasi-planar polygon \cite{Kim99,Lee03}.

\begin{table}
  \caption{Relative energies in meV of the different water cluster anions with respect
  to the D isomers. The dipole moment in Debye, $\mu$, is from the SIC-SCAN 
  calculations of the neutral clusters. The mean absolute deviation (MAD) is with respect to 
  the CCSD(T) values.}
  \label{dEs}
    \begin{tabular}{lcccccccccc}
    \hline
    \hline
    \multirow{2}{*}{Cluster} & \multirow{2}{*}{CCSD(T)$^a$} & \multirow{2}{*}{MP2$^a$} & \multirow{2}{*}{B3LYP$^a$} & \multicolumn{3}{c}{DFA} & \multicolumn{3}{c}{FLOSIC} \\
        & & & & LDA & PBE & SCAN & LDA & PBE & SCAN & $\mu$ \\
    \hline
    3D   &  -- &  -- &  -- &  -- &  -- &  -- &  -- &  -- &  -- & 4.03 \\
    3L   &  69 &  78 &   9 &  99 &  45 &  92 & 145 & 111 & 111 & 7.13 \\
    3AA  & 147 & 173 &  87 & 282 & 131 & 169 & 248 & 156 & 177 & 6.40 \\
    3I-2 & 178 & 204 & 113 & 304 & 144 & 185 & 271 & 174 & 197 & 6.39 \\
    3I-1 & 533 & 551 & 343 & 784 & 407 & 512 & 714 & 461 & 503 & 1.75 \\
    4D   &  -- &  -- &  -- &  -- &  -- &  -- &  -- &  -- &  -- & 5.17 \\
    4AA  & 134 & 178 & 121 & 329 & 185 & 180 & 282 & 208 & 180 & 8.92 \\
    4L   & 191 & 208 & 134 & 375 & 194 & 224 & 317 & 215 & 199 & 9.39 \\
    4I   & 425 & 451 & 308 & 706 & 375 & 429 & 656 & 469 & 471 & 0.00 \\
    5D   &  -- &  -- &  -- &  -- &  -- &  -- &  -- &  -- &  -- & 5.39 \\
    5AA-2& -30 &  30 &  43 &  67 &  84 & -23 & 104 & 149 &  -7 & 9.03 \\
    5AA-1& 139 & 186 & 134 & 305 & 177 & 165 & 294 & 217 &  88 & 9.48 \\
    5L   & 204 & 225 & 147 & 422 & 228 & 242 & 357 & 241 & 225 & 10.77\\
    5I   & 152 & 208 & 139 & 268 & 157 & 127 & 310 & 268 & 165 & 8.71 \\
    6D   &  -- &  -- &  -- &  -- &  -- &  -- &  -- &  -- &  -- & 5.51 \\
    6AA-2& -56 & -13 &  -9 &  21 &  16 & -51 &  70 &  94 &  13 & 10.41\\
    6AA-1& 117 & 182 & 126 & 248 & 151 & 106 & 317 & 291 & 210 & 10.24\\
    6L   & 351 & 356 & 252 & 633 & 327 & 387 & 544 & 345 & 377 & 13.09\\
    6I   & 516 & 533 & 382 & 906 & 446 & 524 & 864 & 596 & 619 & 0.00 \\
    \hline
    MAD  &  -- &  32 &  66 & 383 & 204 &  21 & 168 &  73 &  39 & 0.30 \\
    \hline
    \hline
    {}$^a$ Ref.~\citenum{Yagi08}
  \end{tabular} \\
\end{table}

  In the neutral water dimer, the torsional angle between the bisecting axis of the proton acceptor and the line segment connecting the oxygens is experimentally found to be 57$^o$ \cite{water_dimer_expt}.  \citeauthor{Kim99}  studied several configurations of the water dimer \cite{Kim99} at the CCSD(T) level. They concluded that the presence of the extra electron in the dimer introduces a large change in the torsional angle. They also concluded that the anionic water dimer is a very floppy structure with large vibrational zero-point energy effects. Among the reported structures of the neutral water dimers, the {\it trans} isomer has a dipole moment of 2.7D whereas nearly isoenergetic {\it cis} isomer has a dipole moment of 4.3D.  The individual dipoles of the water molecules are almost parallel in the {\it cis} configuration. Thus, although the total energies of the two anions are very close, it was shown\cite{Kim99} that the {\it trans} isomer barely binds an extra electron, while the {\it cis} isomer has significantly more binding due to its larger dipole.
Based on these earlier reports, FLOSIC calculations were performed on the {\it cis} isomer. In Fig.~\ref{dimer} the EDD of the {\it cis} isomer (2L) and the positions of the FODs are shown. 
As mentioned above, the FODs corresponding to the valence electrons form a tetrahedron 
around each oxygen atom and the extra FOD for the anion finds its optimal place 
at the $\angle$H-O-H bisector, but far away (4.4 \AA) from the acceptor molecule. 
The figure shows the location of the extra electron density in the anion which is in the same direction as the extra FOD.
A similar pattern is observed in all the other clusters. For that reason,
only the extra FOD is shown in figures 2--5, together with the EDD. 

\begin{figure}
    \centering
    \includegraphics[width=0.8\textwidth]{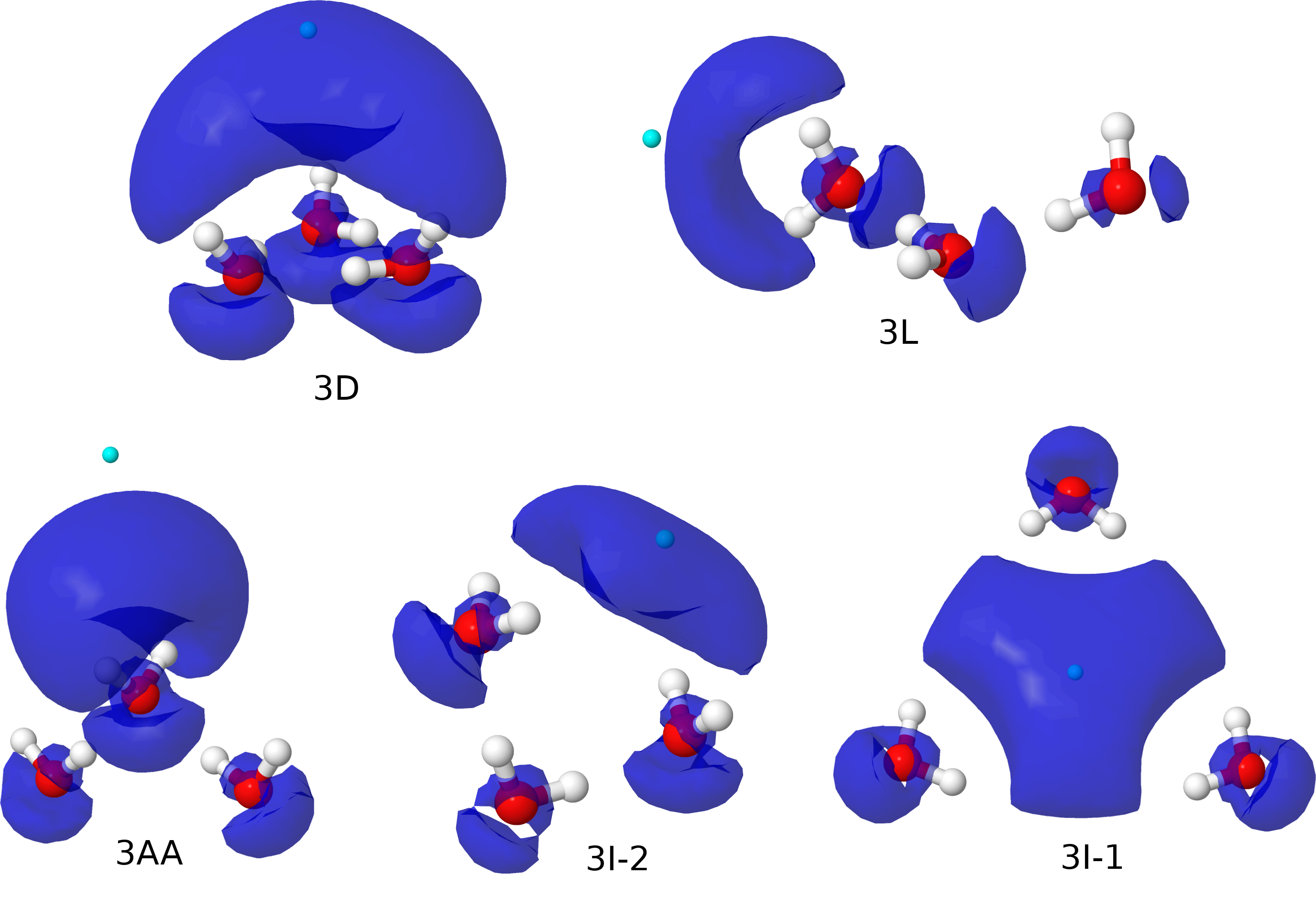}
    \caption{Water trimer anions with the EDD in blue and the optimized extra FOD
    of each isomer.}
    \label{trimers}
\end{figure}

\begin{figure}
    \centering
    \includegraphics[width=0.8\textwidth]{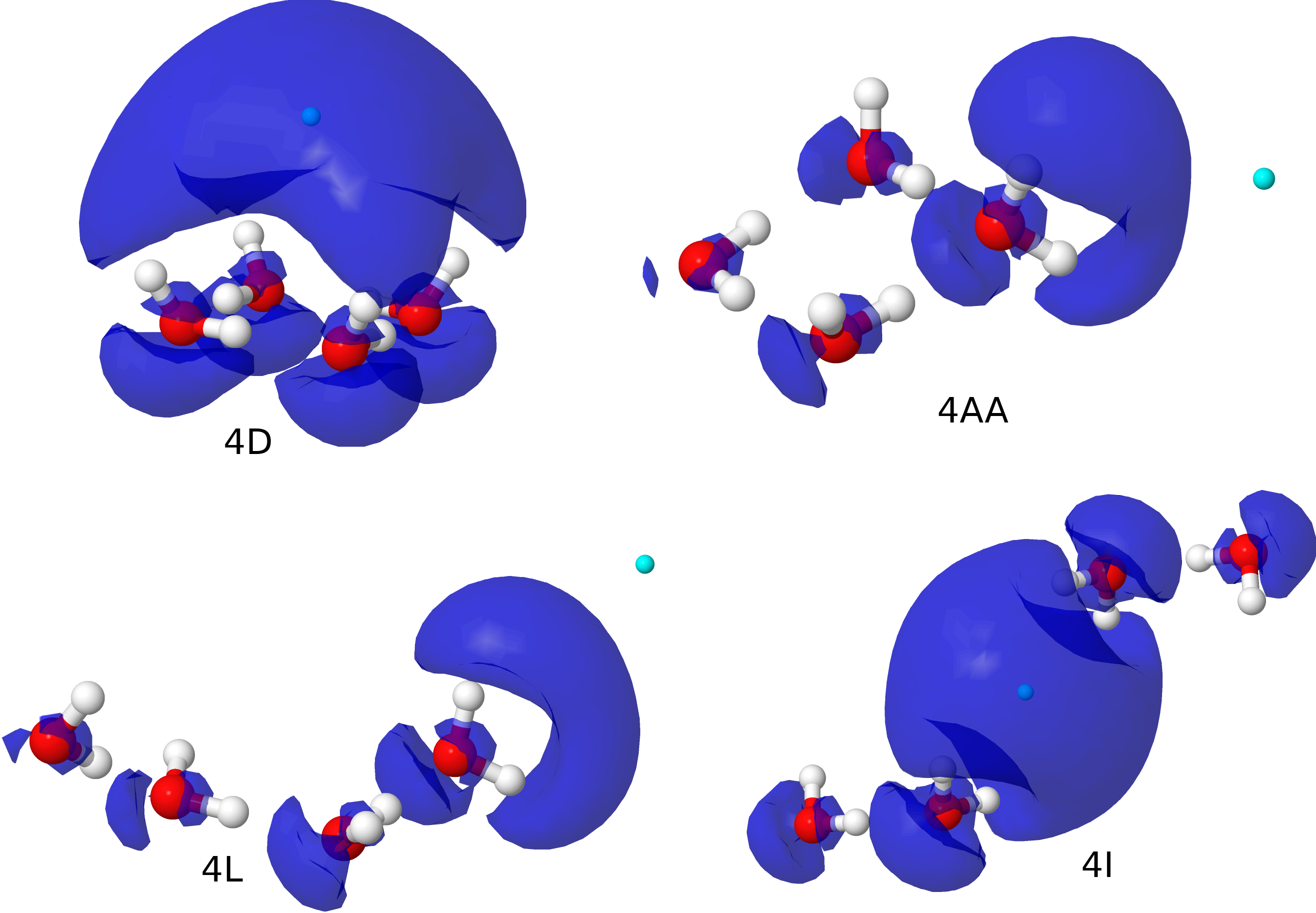}
    \caption{Water tetramer anions with the EDD in blue and the optimized extra FOD
    of each isomer.}
    \label{tetramers}
\end{figure}

\begin{figure}
    \centering
    \includegraphics[width=0.8\textwidth]{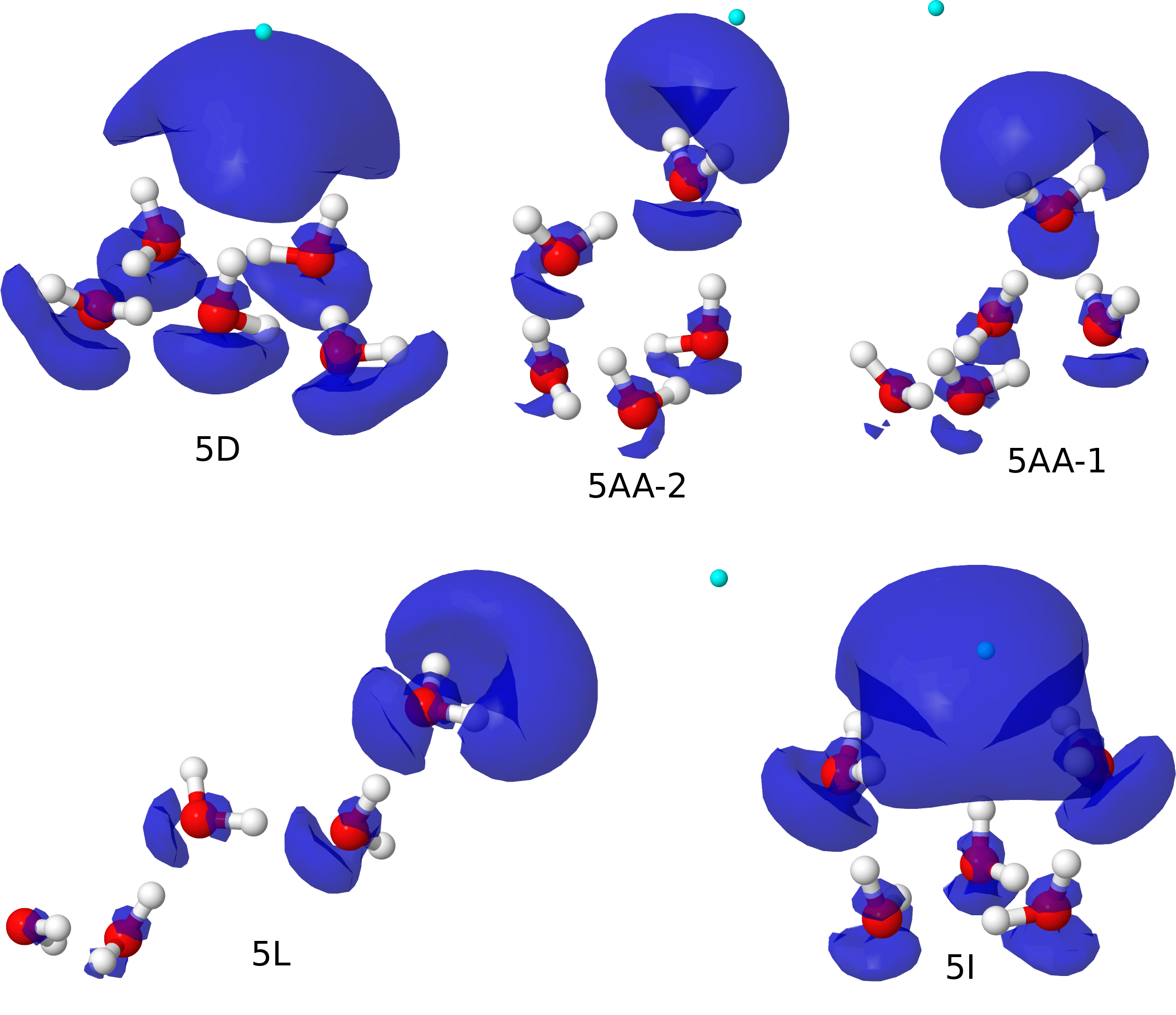}
    \caption{Water pentamer anions with the EDD in blue and the optimized extra FOD
    of each isomer.}
    \label{pentamers}
\end{figure}

\begin{figure}
    \centering
    \includegraphics[width=0.9\textwidth]{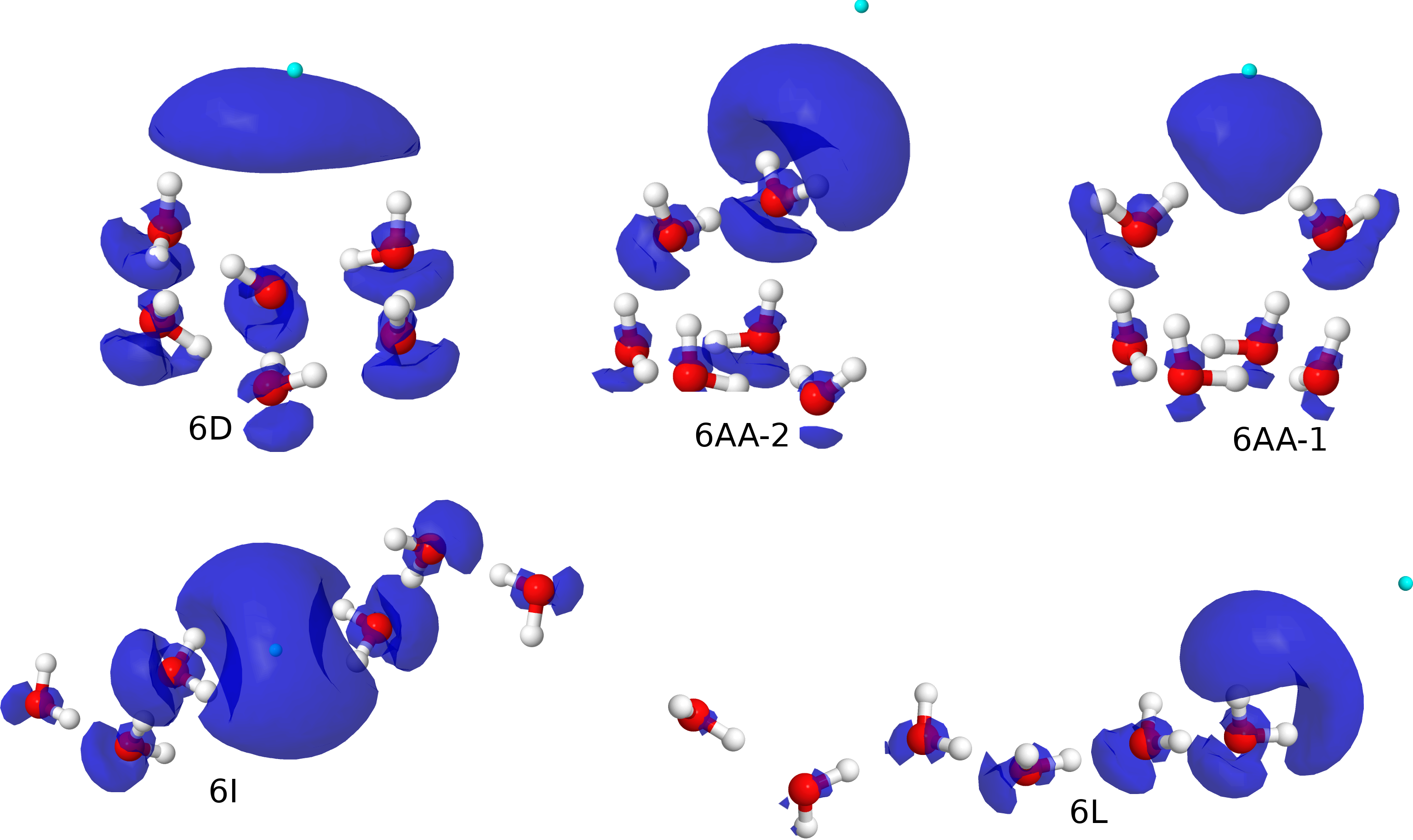}
    \caption{Water hexamer anions with the EDD in blue and the optimized extra FOD
    of each isomer.}
    \label{hexamers}
\end{figure}

The EDD plots of the various trimer, tetramer, pentamer and hexamer isomers are shown in Fig. \ref{trimers} - \ref{hexamers}. The plots corresponding to various functionals are similar and therefore we show only the FLOSIC-LDA version.
The FLOSIC EDD plots are in close agreement with the plots shown in Ref. \cite{Yagi08}. The D structures show a collective cloud of the extra electron spread over multiple water molecules. From the trimer to the pentamer, the D structures have the
water monomers forming a quasi-planar structure, but in case of the hexamers, the monomers in the D structure form a 3-dimensional book structure. However, for all sizes, the excess electron cloud is spread over the whole structure. 
In all the linear isomers, there is one only-donor and one only-acceptor
water molecule with the intermediate molecules being both a donor and an acceptor of a HB.
The excess charge of the anion is mainly accommodated in front of the dangling 
hydrogen atoms of the only-acceptor molecule. On the other hand, the double acceptor isomers have one molecule which attracts the excess electron cloud. In the case of the hexamers, two such double acceptor structures were examined.  In the 6AA-1 isomer,
there are two water molecules that are both double-acceptors.  These two molecules contribute equally to 
accommodate the extra charge in 6AA-1, whereas in the 6AA-2 isomer the extra electron is 
mainly located around the only double acceptor water molecule. 
The I structures with internally trapped electrons are found to be higher in energy for the sizes under study here. The 3I-1 has D3h symmetry and the 4I   structure has a center of inversion and, therefore, both have zero net dipole. Both 3I-2 and 5I structures have two water molecules that form a "bridge" for the extra charge. 

 The EDD plots mainly show the extra electron density as a diffuse cloud. We find that the extra FOD in the anionic cluster generally follows the 
 cloud and is located away from the molecular framework. Our calculations also show that the dipole of the neutral cluster plays a minor role in the binding of the extra charge. As seen from Table \ref{dEs}, the dipole moment of the lower-energy clusters are in fact smaller than those of many high-lying isomers. In general, the linear (L) structure, which has a chain of HBs, has a larger dipole moment compared to the ring-like D structure. 

The relative energies of the anionic water clusters with n=3-6 are presented in Table \ref{dEs}. For the sizes n=3 and 4, the relative energies follow similar trends across all methods used here, including the FLOSIC-DFAs.  The D structures of the trimer and tetramer are found to be the lowest energy isomer by all the methods (Cf. Table \ref{dEs}). The D structures are ring-like for both sizes. For the trimers, all the computational methods also put the linear isomer (3L) as the second lowest isomer. However, Hammer et al.\cite{Hammer05a} identified the linear structure for the trimer anion from infrared spectrum  and theoretical calculations on the vibrational spectra of the trimer anion. Similarly, the 4AA isomer 
has been experimentally identified for the tetramer. \cite{Shin04}. 
\citeauthor{Shin04} also found weaker signals in their photo-electron spectra that can be ascribed to 4L and 4D,
but the major presence of the 4AA isomer was unequivocally confirmed through IR 
experiments and B3LYP vibrational spectra comparisons \cite{Hammer05}.The 5D isomer, which is
presumably optimized from the $C_5$ ring with the five oxygen atoms forming a
planar pentagon, does not conform to a planar structure upon optimization. 
One water molecule moves out-of-plane,
while the two adjacent molecules reorient their dangling hydrogen atoms radially
outward thus diminishing the characteristic collective cloud of the D isomers. From the pentamer onward, the D structure is not the lowest energy isomer at the CCSD(T) level. The same ordering is also seen with the SCAN functional. 

Since the first experiments on the water clusters, the hexamer was identified as a magic number for
the water cluster anions \cite{Armbruster81,Coe90} and has been extensively studied.
It is broadly accepted that the experimental signal comes mainly from the 6AA-2 isomer 
with some contribution of the 6D\cite{Bailey96,Lee03,Hammer05b}. 
The CCSD(T) as well as MP2 and B3LYP calculations from Ref. \cite{Yagi08} predict the 6AA-2 structure as the lowest energy isomer. The SCAN meta-GGA functional makes the same
prediction. On the other hand, the FLOSIC method with all  three different functionals predicts the 6D isomer to be the lowest energy structure.
The 6D is the so-called book
structure and the collective cloud is more concentrated on the two upper molecules as depicted in Fig.~\ref{hexamers}. 
The 6AA-2 can be thought of as
a combination of the 4D isomer and the water dimer. The 6AA-1 and 6AA-2 structures are
related to the so-called prism structure, which is one of the most stable neutral hexamers.
The difference is that in the structure of the anions, the prism has one broken edge.
In the 6AA-1 isomer there are two double acceptor molecules without any HB between them. These two molecules contribute equally to 
accommodate the extra charge in 6AA-1, whereas in the 6AA-2 isomer the extra electron is 
mainly located around the single double acceptor water molecule. 6AA-2 is favored over 6AA-1 by CCSD(T), MP2, B3LYP, and the SCAN functional. The relative energies of these double acceptor clusters are comparable in SCAN and CCSD(T).
The 6L and 6I isomers, with the largest and  smallest dipole moments, respectively, have significantly higher energies in all methods and are not likely to be found experimentally. Although the isomer ordering of the hexamers using SCAN agrees with that of CCSD(T), the SIC-SCAN results show same ordering as for SIC-LSDA and SIC-PBE.

 Overall, the SCAN meta-GGA functional results for the stability of the anionic clusters are in close agreement with the CCSD(T) predictions. The ordering of all the isomers is the same  and the mean absolute deviation (MAD) of relative energies is only 21 meV. The energy ordering of isomers for the smaller anionic clusters containing 3 and 4 water molecules is the same for all three functionals and their FLOSIC counterparts. For  n=5 and 6, LDA and PBE and FLOSIC-LDA and FLOSIC-PBE all invert the ordering of the two lowest-energy isomers.  Both SCAN and FLOSIC-SCAN give the correct order for these isomers for n = 5, but only SCAN does for n = 6.  We also find that the deviation of the relative energies from the CCSD(T) values is reduced significantly upon
 removal of self-interaction error from LDA and PBE.  For LDA, the MAD decreases from 383 meV to 168 meV, and for PBE,  from 204 meV to 73 meV.  On the other hand, correcting for self-interaction in SCAN increases
 the MAD from 21 meV to 39 meV. The performance of the SCAN functional is
 the best among all approximations including MP2.  This observation is consistent with earlier results where the SCAN functional was found to provide an excellent description of the ordering\cite{SCAN15,waterSIC} of the neutral water hexamers,  as well as the ordering of various phases of ice. It should be noted
  that SCAN is a meta-GGA functonal whose correlation component is already self-interaction free.
  Removal of SIE from SCAN results in over-correcting, resulting in a somewhat
  degraded performance. However, FLOSIC-SCAN results are still comparable to MP2.

\subsection{Vertical detachment energies}
The VDEs calculated from total energy differences between neutral and anionic clusters with and without FLOSIC are presented in Table \ref{VDEs}. 
Accurate description of the VDE is a challenge for the density functional
approximations due to the inherent self-interaction present in these DFAs. 
Presence of SIE, in general, results in 
excessive electron delocalization causing significant errors in the electron binding 
energies. Indeed, as can be seen from the Table \ref{VDEs}, 
the errors made by the pure DFAs are substantial. Since the CCSD(T) values are in excellent agreement with the available experimental values and also because the CCSD(T) values are available for all the clusters studied here, we calculate the errors in DFAs and SIC-DFAs with respect to the CCSD(T) results.  The mean absolute error (MAE) 
in VDE with respect to CCSD(T) results are 339, 247, and 127 meV for LSDA, PBE, and
SCAN, respectively. These numbers highlight the tendency  
of these functionals to over-bind an extra electron to water clusters, but they also show that as one climbs the Perdew-Schmidt Jacob's
ladder\cite{perdew2001jacob} of increasingly complex functionals, 
the tendency decreases. The
fourth rung of functionals corresponds to the hyper-GGA functionals. The hybrid
functionals that include a certain fraction of HF exchange belong to this rung.
In many situations, HF calculations have errors that are opposite to those made by DFAs. This often results in better descriptions of properties by hybrid 
functionals compared to standard
DFAs. However, in the case of 
the VDE of water cluster anions, the popular global hybrid B3LYP gives no improvement. In fact, it can be seen from Table \ref{VDEs} that the MAE
with B3LYP (238 meV) is comparable to that with PBE,
which sits two rungs lower on Jacob's ladder. The performance of 
SCAN is substantially
better than B3LYP . This may be due to the correlation component of the
SCAN functional being self-interaction free. It is consistent with recent reports\cite{SCAN15,Chen17} that show SCAN provides a much improved description of 
liquid water and neutral water clusters than other DFAs, and with the fact that SCAN gives the correct ordering of water cluster anion isomers ordering as discussed 
above.  

The relatively poor performance of B3LYP despite having an admixture of HF exchange is 
perplexing. 
The over-binding tendency of B3LYP, or, in other words, the effect 
of  mixing HF with DFAs on the electronic
properties, was investigated by Yagi
and coworkers\cite{Yagi08}. These authors studied water cluster anions using
the long-range corrected DFAs (LC-DFAs). In this approximation,
the electron repulsion operator is divided into short-range and long
range parts with the short range part being described by the DFA and the long-range orbital-orbital exchange interaction described using HF
exchange. They used the BLYP \cite{Becke88,LYP88} and BOP\cite{Becke88,Tsuneda99} functionals 
for the short-range. Their 
results showed that both the LC-BLYP and LC-BOP functionals provided significant 
improvement in predicting VDEs over the global hybrid B3LYP.  They attributed
the excellent performance of the LC-BOP functional (MAE $~$ 35 meV) over 
the LC-BLYP functional (MAE $~$ 152 meV) to the satisfaction of 
fundamental conditions by the BOP functional and violation
of them by BLYP.
Their results show that good estimates of VDE comparable to those of MP2 can be obtained using
long range corrected functionals. 

It is interesting to compare the 
performance of these range-corrected functionals with the present approach where the 
self-interaction is explicitly removed on an orbital by orbital basis 
using the Perdew-Zunger method. We find that the removal of self-interaction error
significantly improves the VDE obtained from the total energy differences of anion and neutral clusters. The 
VDEs obtained from the total energy differences for the SIC-LSDA, SIC-PBE, and SIC-SCAN functionals  are summarized in Table \ref{VDEs}. The MAE in VDEs for the three SIC-DFAs are 180 meV for SIC-LSDA,
60 for SIC-PBE, and 57 meV for SIC-SCAN. These errors can be compared to 
those for the uncorrected functionals -  339 meV for LSDA, 247 meV for PBE, and
127 for the SCAN. This indicates that the SIE is a large contributor 
to the over-binding tendency of these approximations.

To better understand the nature of these errors, we 
used the self-interaction corrected densities obtained in the SIC-DFA calculations 
to evaluate the uncorrected DFAs. We then used these to recompute the VDEs.  Following the notation adopted in 
the literature, this approach is called DFA@SIC-DFA\cite{verma2012increasing}. Using this scheme, 
(Cf. Table \ref{atVDEs}) we
find that the MAE in LSDA@SIC-LSDA, PBE@SIC-PBE, and SCAN@SIC-SCAN to be 
303, 217, and 110 meV, respectively. This is only a slight (< 15\%) improvement over the self-consistent DFA values.  The primary 
source of the errors in VDEs therefore is the approximate functional and not simply the density.
Burke and coworkers\cite{kim2013understanding} categorized DFT calculations as either normal 
or abnormal depending on whether errors stem from the approximate functional (normal) or from the approximate density (abnormal). The comparison of VDES from the DFAs, and DFAs@SIC-DFAs show that the VDE calculations are apparently normal.

One drawback of the pure DFAs (LSDA, PBE, and SCAN) discussed above is that
although the total energy of the water cluster is correctly lowered with the addition of an
extra electron, the eigenvalue of the extra electron 
is positive within these approximations. 
The over-binding tendency of the DFAs discussed earlier
is for VDE estimates made from the total energy difference between corresponding anion and neutral
clusters. In exact DFT, the highest occupied eigenvalue equals the negative of the ionization 
potential\cite{perdew1982density,levy1984exact,perdew1997comment,PhysRevB.60.4545}. This relationship does not strictly
hold for approximate density functionals and in most DFAs, the absolute value of the HOMO eigenvalue substantially
underestimates the first ionization potential due to self-interaction error.
The positive eigenvalue of the extra electron in the water anions in DFT indicates that the extra electron 
is not actually bound in the complete basis set limit. The positive eigenvalue is 
a result of self-interaction error 
which makes the asymptotic potential seen by the electron positive. 
Removing self-interaction
improves the asymptotic description of the potential and results in negative
(bound) eigenvalues for the extra electron in all three SIC-DFAs 
used in this work.
The improved description of the binding of the extra electron due to self-interaction correction can be seen from Table \ref{VDEs}
which presents predictions of the 
VDE from the eigenvalues of the extra electron in the SIC-DFA calculations (VDE = $- \epsilon_{HOMO}$). 
The eigenvalues are excellent approximations to the VDEs, especially
for SIC-PBE, for which the MAE with respect to CCSD(T) estimates is only 17 meV. The MAE
for SIC-SCAN and SIC-LSDA eigenvalues are 44 and 117 meV, respectively. It is not 
obvious why the SIC-PBE eigenvalues approximate CCSD(T) VDE better than SIC-SCAN or 
SIC-LSDA. 
The SIC-SCAN HOMO eigenvalues agree better with the available experimental VDE values, with a MUE of 29 meV compared to 35 meV for SIC-PBE. The SIC-LDA eigenvalues have a MUE of 70 meV when compared with the experimental values.  For all three functionals, the SIC-DFA HOMO eigenvalues are better approximations of the VDEs than the total energy differences. The shift of the anion eigenvalues to positions close to the removal energies underscores that 
the self-interaction correction is needed for a more physically correct description of water cluster anions with DFAs.

It should be noted that the positive HOMO eigenvalues in the uncorrected DFA calculations imply that the over-binding of the VDE by DFA total energy differences is actually worse than seen in Table \ref{VDEs}.  The energy of the cluster anion  would be lowered by removing a fraction of the extra electron to a large distance from the cluster.  The minimum energy state corresponds to removing sufficient charge to make the HOMO eigenvalue zero.  Lower anion energies would give still large VDEs than those in Table \ref{VDEs}.

\begin{table}
 \rotatebox{90}{
  \caption{Vertical detachment energies and HOMO eigenvalues in meV of water cluster anions. 
  The mean absolute deviation (MAD) is respect to the CCSD(T) values.
  The uncertainties of the experimental values are about $\pm 30$ meV.}
  \label{VDEs}
  \begin{tabular}{|lccccc|ccc|ccc|ccc|}
    \hline
    \multirow{2}{*}{Cluster} & \multirow{2}{*}{CCSD(T)$^a$} & \multirow{2}{*}{MP2$^a$} & \multirow{2}{*}{B3LYP$^a$} & \multirow{2}{*}{LC-BOP$^a$} & \multirow{2}{*}{Expt} & \multicolumn{3}{|c}{DFA-VDE} & \multicolumn{3}{|c}{SIC-VDE} &
    \multicolumn{3}{|c|}{SIC-HOMO} \\
        & & & & & & LDA & PBE & SCAN & LDA & PBE & SCAN & LDA & PBE & SCAN \\
    \hline
    2L    &  29 &   9 & 194 &  28 & $50^b$& 240 & 205 &  96 & 121 &  77 &  63 &  75 &  30 &  42 \\
    3D    &   6 &$-14$& 184 & $-9$&       & 216 & 184 &  67 &  96 &  58 &  39 &  53 &  14 &  22 \\
    3L    & 146 & 115 & 346 & 161 &$130^b$& 430 & 373 & 258 & 288 & 211 & 203 & 229 & 148 & 177 \\
    3AA   & 187 & 146 & 399 & 202 &       & 489 & 411 & 300 & 348 & 245 & 236 & 276 & 180 & 216 \\
    3I-2  & 175 & 138 & 427 & 198 &       & 488 & 413 & 293 & 348 & 251 & 225 & 279 & 185 & 215 \\
    3I-1  & 190 & 155 & 526 & 227 &       & 589 & 502 & 354 & 438 & 310 & 233 & 346 & 205 & 244 \\
    4D    &  49 &  22 & 239 &  21 & $60^c$& 289 & 244 & 123 & 158 & 105 &  69 & 109 &  62 &  73 \\
    4AA   & 336 & 283 & 561 & 273 &$350^c$& 678 & 577 & 469 & 511 & 386 & 390 & 440 & 314 & 375 \\
    4L    & 255 & 214 & 478 & 283 &$250^c$& 558 & 483 & 362 & 407 & 307 & 313 & 333 & 236 & 276 \\
    4I    & 439 & 394 & 713 & 489 &       & 861 & 729 & 610 & 688 & 494 & 492 & 642 & 456 & 524 \\
    5D    &  61 &  31 & 285 &  18 &       & 329 & 279 & 146 & 184 & 130 & 103 & 129 &  73 &  89 \\
    5AA-2 & 376 & 318 & 592 & 408 &$410^b$& 732 & 621 & 511 & 561 & 418 & 433 & 484 & 354 & 415 \\
    5AA-1 & 370 & 313 & 600 & 366 &       & 738 & 625 & 510 & 574 & 437 & 432 & 501 & 358 & 407 \\
    5L    & 294 & 250 & 527 & 338 &       & 611 & 532 & 407 & 442 & 354 & 348 & 373 & 277 & 322 \\
    5I    & 469 & 406 & 757 & 516 &       & 891 & 754 & 637 & 730 & 539 & 587 & 644 & 467 & 549 \\
    6D    & 104 &  63 & 347 &  58 &$210^b$& 413 & 342 & 205 & 255 & 187 & 164 & 188 & 114 & 132 \\
    6AA-2 & 477 & 414 & 706 & 507 &$480^b$& 856 & 728 & 621 & 684 & 513 & 547 & 598 & 443 & 508 \\
    6AA-1 & 553 & 482 & 847 & 634 &       &1026 & 865 & 746 & 814 & 600 & 653 & 737 & 521 & 616 \\
    6L    & 381 & 331 & 643 & 426 &       & 729 & 636 & 505 & 552 & 445 & 436 & 474 & 358 & 402 \\
    6I    & 839 & 793 &1120 & 922 &       &1349 &1163 &1047 &1131 & 870 & 915 &1145 & 904 &1001 \\
    \hline
    MAD   &  -- &  44 & 238 &  35 &       & 339 & 247 & 127 & 180 &  60 &  57 & 116 &  17 &  44 \\
    \hline
    \multicolumn{4}{l}{{}$^a$ Ref.~\citenum{Yagi08}, {}$^b$ Ref.~\citenum{Kim98}, {}$^c$ Ref.~\citenum{Shin04}}
  \end{tabular}
 }
\end{table}
\begin{table}
  \caption{VDEs (in meV) obtained using DFA total energies computed with self-consistent FLOSIC-DFA densities (DFA@FLOSIC) compared 
  to the reported VDEs obtained with CCSD(T). 
  The mean absolute deviation (MAD) is respect to the CCSD(T) values.}
  \label{atVDEs}
  \begin{tabular}{lcccc}
    \hline
    \hline
    Cluster  & LDA & PBE &SCAN &CCSD(T)$^a$ \\
    \hline
    2L   & 217 & 184 &  82 &  29 \\
    3D   & 191 & 165 &  58 &   6 \\
    3L   & 403 & 303 & 237 & 146 \\
    3AA  & 456 & 383 & 280 & 187 \\
    3I-2 & 389 & 389 & 275 & 175 \\
    3I-1 & 467 & 468 & 327 & 190 \\
    4D   & 264 & 223 & 111 &  49 \\
    4AA  & 651 & 545 & 446 & 336 \\
    4L   & 532 & 453 & 340 & 255 \\
    4I   & 844 & 704 & 584 & 439 \\
    5D   & 299 & 248 & 114 &  61 \\
    5AA-2& 712 & 586 & 547 & 376 \\
    5AA-1& 699 & 593 & 493 & 370 \\
    5L   & 581 & 496 & 386 & 294 \\
    5I   & 873 & 725 & 627 & 469 \\
    6D   & 380 & 311 & 190 & 104 \\
    6AA-2& 832 & 697 & 601 & 477 \\
    6AA-1& 983 & 808 & 720 & 553 \\
    6L   & 695 & 601 & 484 & 381 \\
    6I   &1331 &1142 &1024 & 839 \\
    \hline
    MAD  & 303 & 217 & 110 &  -- \\
    \hline
    \hline
    {}$^a$ Ref.~\citenum{Yagi08}
  \end{tabular}
\end{table}

\section{Conclusion}
 We have used the recently developed Fermi-L\"owdin orbital self-interaction correction scheme with the LSDA, PBE, and SCAN meta-GGA functionals 
 to study small water cluster anions. Our results show that the SCAN functional provides a very good description
of isomer ordering, as well as the relative energies of isomers, when compared to CCSD(T) results. The application of FLOSIC significantly improves the agreement for SIC-LSDA and SIC-PBE relative to CCSD(T), however, the SIC-SCAN results deviate somewhat more from the reference values than SCAN.
The excellent performance of SCAN for binding energies does not carry over to the description of the binding of the extra electron. The 
SCAN MAE (127 meV) for electron detachment energy, although smallest among
the LSDA (339 meV),
PBE GGA (247 meV) and also earlier reported B3LYP (238 meV) results, 
is still substantially larger than the MP2 MAE (44 meV). 
Removing self-interaction results in significantly improved VDEs for all functionals, with about 60 meV 
errors for the SIC-SCAN and SIC-PBE.  Similarly, removing self-interaction
is essential for obtaining orbital energies that are consistent with electron binding.  For SIC-PBE, the HOMO eigenvalues give remarkably good predictions of VDEs, with a MAE with respect to CCSD(T) of only 17 meV.
An interesting feature of the FLOSIC calculations is the chemical insight that can be gained from the Fermi-orbital descriptor (FOD) positions. The FOD associated with the extra electron indicates where the
the excess charge is accommodated in the clusters. In several cases, the FOD position 
is relatively far away from the cluster center, indicating a more delocalized density for the extra electon.

\begin{acknowledgement}
The authors acknowledge useful discussions with Dr. Luis Basurto. This work was supported by the US Department of Energy, Office of 
Science, Office of Basic Energy Sciences, as part of the 
Computational Chemical Sciences Program under Award No. 
DE-SC0018331. The work of R.R.Z. was supported in part by
the US Department of Energy, Office of Science, 
Office of Basic Energy Sciences, under Award No. DE-SC0006818.
Support for computational time at the Texas Advanced 
Computing Center through NSF Grant No. TG-DMR090071, 
and at NERSC is gratefully acknowledged.
\end{acknowledgement}

\bibliography{SICpapers,refs,refs2}

\end{document}